# Drowsy Driver Detection by EEG Analysis Using Fast Fourier Transform

Mejdi Ben Dkhil, Ali Wali, and Adel M. Alimi
*REsearch Groups in Intelligent Machines University of
Sfax, National School of Engineers (ENIS) BP 1173,
3038
Sfax, Tunisia*
{mejdi.bendkhil, ali.wali, adel.alimi}@ieee.org

*Abstract*— In this paper, we try to analyze drowsiness which is a major factor in many traffic accidents due to the clear decline in the attention and recognition of danger drivers. The object of this work is to develop an automatic method to evaluate the drowsiness stage by analysis of EEG signals records. The absolute band power of the EEG signal was computed by taking the Fast Fourier Transform (FFT) of the time series signal. Finally, the algorithm developed in this work has been improved on eight samples from the Physionet sleep-EDF database.

## I. INTRODUCTION

In literature, we find four types of drowsiness: sleepiness, hardly force, period of day and physical annoying. Extremely engaged to finish their works along the day, persons do not possess occasion to obtain sufficient relax. Since an experiment to hold attentive, they can usually camp to caffeine or another energizers. Thus, standing attentive longer than one day, the person may not be apt to gate it more which will provoke its failure and abnormally affect the body. In addition, the body brain is recorded form on epochs of day circumstance. Indeed, it controls the person and monitors the epochs when it must be dormant or be attentive. Usually, these epochs will be influenced by downing and darkness. Hence, continuing the epoch attentive negatively influences the person [3]. In addition, the physical aspect is interested; we can affirm that we find persons who have physical sickness that can conduct to drowsiness position. One more, being anxious or starved may conduct the driver to land drowsiness state [4].

It is common to say that EEG is the most important and useful technique used to measure and capture the brain signals because of its great temporal resolution, noninvasiveness, and low set-up costs. In fact, your brain is the boss of your body. It runs the show and controls just about everything you do, what you think and feel, how you learn and remember, and the way you move and talk, the beating of your heart, the digestion of your food, and yes, even the amount of stress you feel. Then the nervous system is like a network that relays messages back and forth from the brain to different parts of the body and governs the activities of different organs of the body.

### A. Electroencephalographic signals

As the characteristic patterns of the electrical potentials can change from one state to another. Considering everything they do, EEG helps to show in what state the person is whether s/he is happy, nervous, even asleep… epilepsy and brain computer interface (BCI) are mostly the important parts of EEG. Seizure can be defined as the transient of weird and unusual behavior of neurons within one or several neural networks, which restricts the patients' physical and mental activities. Yet, EEG has a focal role in nervous electro-physiology area such as using spike wave in order to discover diagnose epilepsy, brain tumor early, sleep analysis and monitor the depth of anesthesia etc.

In literature, brain signals are generally employed to evaluate drowsiness is termed the Electroencephalogram. This signal possesses various frequency waves, containing the delta band which is associated to sleep position, the theta band which describes to drowsiness. Relaxation is defined by the alpha band and the alertness is described by the beta band. The decrement of the alpha band and the increment in the theta frequency, express drowsiness [5].

TABLE I. EEG PULSES

| Rhythms | Frequency interval | Location | Reason |
|---|---|---|---|
| Delta | (0-4) Hz | Frontal lobe | Deep sleep |
| Theta | (4-7) Hz | Median, temporal | Drowsiness and meditation |
| Alpha | (8-13) Hz | Frontal, occipital | Relaxation and closed eyes |
| Mu | (8-12) Hz | Central | Controlateral and motor acts |
| Beta | (13-30) Hz | Frontal, central | Concentration and reflection |
| Gamma | (>30) Hz | — | Cognitive functions |

By Analyzing of Table I, we remarks that The Theta activity is concerned for drowsy driving recognition and the Alpha activity is prevailing when a person is snoozing or closing his eyes. All along the transition from awake to sleep or drowsy state; drowsiness instructions can be sent by a wide-range of multimodal data signals. For drowsiness detection for Human Computer Interaction (HCI), the signals may be divided into four types.

*B. Brain computer interface (BCI)*

It is a system that lets people get across with the external world only by thinking without depends on muscular or nervous activity [6]. BCI starting by controlling the user's brain activity which is transmitted into brain signals braced to obtain characteristics collected into a vector termed the "feature vector". The aligning of the concluding results into commands to be used by the system that shows a feedback to the user in order to calibrate or adjust their brain activity. In order to calculate brain expressions, diverse realizing technologies have been imported such as the Functional Magnetic Resonance Imaging (FMRI), Positron Emission Tomography (PET), and Functional Near-Infrared Brain Monitoring (FNIBM). Because of its high time resolution, noninvasiveness, ease of acquisition, and cost effectiveness, the EEG is the adequate brain controlling technique in current BCIs.

*C. Related Work*

Through this part, we are focusing on recent approaches to calculate the drowsiness's level, if car technologies are going to secure or at least alert of driver drowsiness, what expressions does the conductor give off that can be mentioned? In literature, diverse categories of technologies In literature, diverse categories of technologies that offer driver drowsiness detection. Those techniques can be partite into three main types:

- Vehicle based measures: A diverse metrics which contains detours from lane position, change of the steering wheel, tension on the acceleration pedal, etc., are regularly monitored. In spite of, any variation in these standards adduces to the possibility of a significant increment in the driver drowsiness level.

- Behavioral based measures: We use a camera to monitor the driver's behavior (yawning, eye blinking, textures, etc) [7], and the driver will be warned if any of these drowsiness signs are observed.

- Physiological based measures: Drowsiness level is measured by detecting heart beat, brain information and pulse rate.

In reality, drowsiness detection is depended on the three parameters indicated overhead. In recent work, these measures should be conducted to maintain observation on the present approaches, issues correlated with them and the improvements that must be mentioned to produce a robust technique.

In this paper, the proposed system consists of EEG analysis

## II. PRESENT WORK

*A. System Overview*

In this part, we used an EEG headset in goal to evaluate the drowsiness state. To collect the EEG Data, we include used the Emotiv EPOC headset containing 14 electrodes (AF3, F7, F3, FC5, T7, P7, O1, O2, P8, T8, FC6, F4, F8 and AF4) and 2 references electrodes. These electrodes are located in relation to the standard design of the 10-20 system. This system guides to the architecture of the electrodes changing from 10 or 20% depending to the morphology of the individual (Fig.1.).

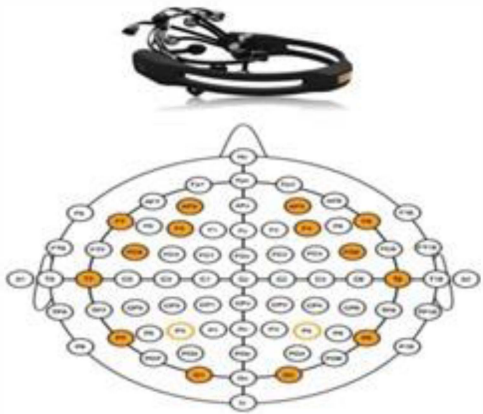

(1.a) Emotiv EPOC headset

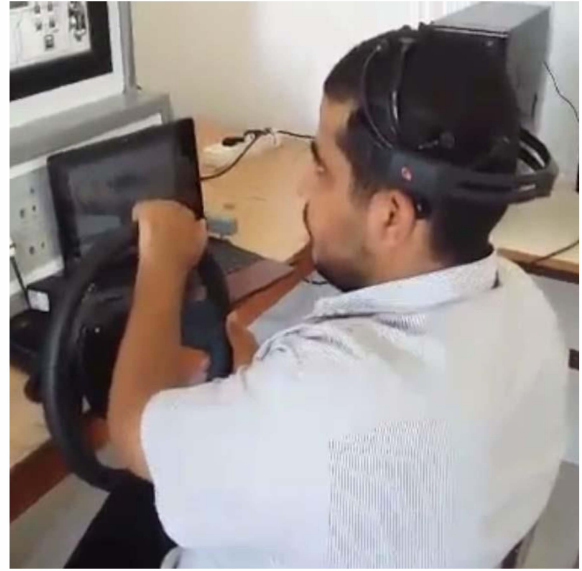

(1.c) Driving car environment

Fig.1. System overview

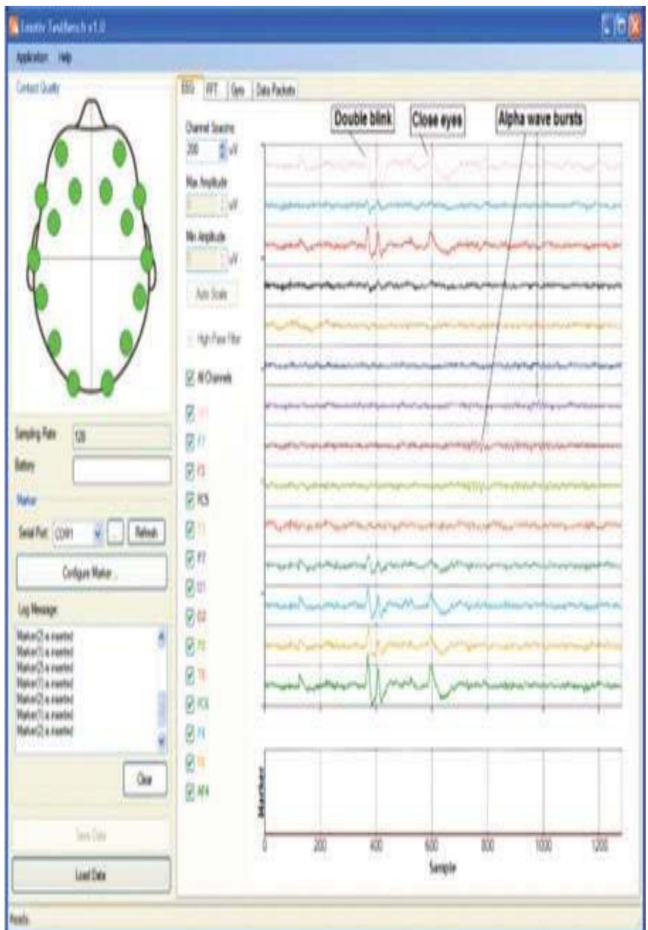

(1.b) Electors positions and signals

### B. System Flowchart

In this part, we present the architecture of the proposed system.

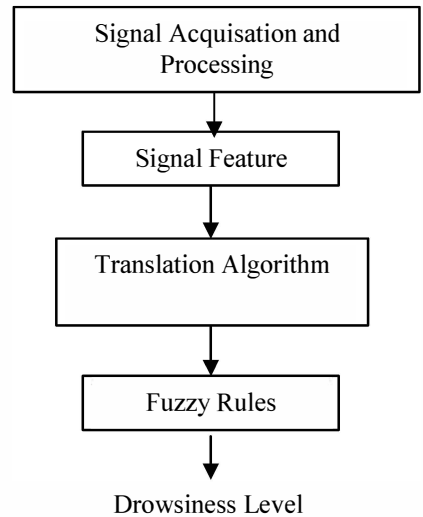

Fig.2 Approach for proposed driver drowsiness detecting system [8].

### C. FFT and Pass-band filtering

For the extraction phase, some researches applied the wavelet [9], and some others work with FFT. In our work, we have considered an EEG signal of a period of 20 seconds that may be asset in the starting, in the middle or at the closing of the experience. Basically, we have adapted the FFT on the signal, and then a pass-band filtering is used to obtain the expected frequency bands.

We define 6 principles EEG pulses: delta, theta, alpha, mu, beta and gamma. Every one of them is further remarkable in various positions as represented in the Table I: In proportion for our objectives, we propose to use only the Alpha and Beta pulses that are the best particular pulses in the drowsiness state inspection. To follow the measures of these pulses in six positions in the frontal lobe: AF3, AF4, F3, F4, FC6 et F8, and a one situation in the parietal lobe: P8.

### D. Feature extraction

- Arousal (A): A large beta capacity and a low alpha correspondence in the parietal lobe. Beta pulse is relied on state of a warning or awakened mind, although alpha wave is particular leading in a case of resting. Thus, the beta/alpha ratio is a moderate character of the activity case of a body [10].

$$A = \frac{[\alpha(AF3 + AF4 + F3 + F4)]}{[\beta(AF3 + AF4 + F3 + F4)]} \quad (1)$$

- Valence (V): The prefrontal lobe (F3 and F4) shows a deciding character in the performance of drowsiness and awakened experience.

$$V = \frac{\alpha F4}{\beta F4} - \frac{\alpha F3}{\beta F3} \quad (2)$$

- Dominance (D): In the frontal lobe we remark a rise in the ratio beta / alpha and in the parietal we conclude a rise in the beta power.

$$D = (\beta FC6/\alpha FC6) + (\beta F8/\alpha F8) + (\beta P8/\alpha P8) \quad (3)$$

Through obtain the arousal, valence and dominance characteristics; we propose the coming method [11]:

a) Charging an XLS file.

b) For every one of data of the signal: (treatment of 20 s of the signal).
• Adapting the FFT filter.

• Creation of the two pass-band filter Alpha and Beta.

• Adapting the pass-band filter on the different electrodes.

• Estimating the rates of Arousal, valence and dominance.

    End
    For.
c) Clustering by Fuzzy Cmeans.

### E. Fuzzy logic classification

For the classification of the different pulses, we adopt the Mamdani fuzzy logic technique (Fig.3).

Arousal (A), valence (V), dominance (D) is three inputs and the state of drowsiness (DS) present the output variable.

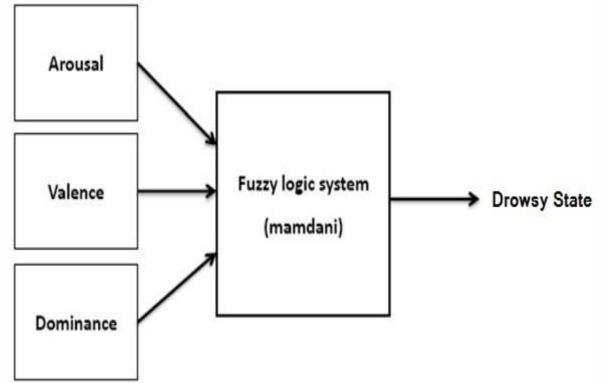

Fig.3. Fuzzy Logic System

To evaluate the drowsiness state, we go to define three membership functions for each input variable: Small (S), Medium (M) and Large (L).

In this stage, to evaluate the DS refers to propose nine fuzzy rules:

1) if (A = M) then (D = S)

2) if [(A = S) and (V = S) and (D = S)] then (DS = S)

3) if [(A = L) and (V = L) and (D = L) ] then (DS = L)

4) if [(A = L) and (V = S) and (D = M)] then (DS = S)

5) if [(A = L) and (V = S) and (D = L)] then (DS = S)

6) if [(A = S) and (V = M) and (D = M)] then (DS = S)

7) if [(A = S) and (V = L) and (D = M)] then (DS = M)

8) if [(A = S) and (V = M) and (D = L)] then (DS = S)

9) if [(A = S) and (V = L) and (D = L)] then (DS = M)

### F. Analysis Process

To validate of the chosen classifier, the Mamdani call the "Fuzzy logic controller" contained in the Simulink library. In this function, the Simulink bloc of the EEG pulses evaluates contains a Matlab function (Function1). The task of Function1 is to recall the file including the different pulses, comply with the FFT on this pulse, use the passband filtering to obtain the different pulses as well as we deduce the characteristics of the signal by designing the three inputs ( A, V and D) rates.

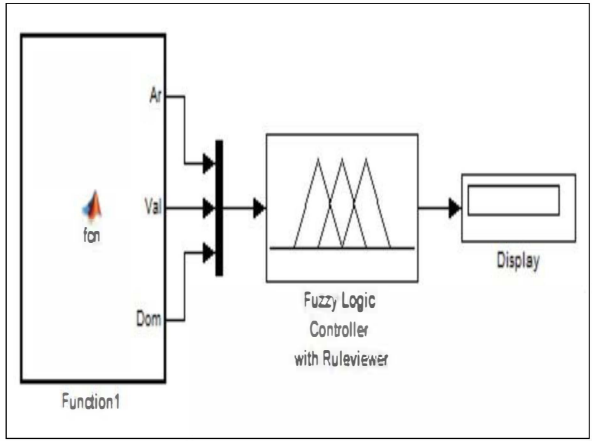

Fig. 4. Simulink bloc of EEG signals analysis

By analyzing of the Fig.4, we conclude that valence, arousal and dominance are the inputs to the fuzzy logic controller that decide to classify the characteristics of the created signal and produce the drowsiness level of the experiment sample.

### III. EXPERIMENTS RESULTS

In this phase, we explain diverse experimental results which are presented in this paper.

#### A. Database

We are going to validate our proposed technique on eight samples from the Physionet sleep-EDF database [12]. The files are in EDF/EDF+ (European Data Format). Fig. 5 illustrates the EEG data from one sample.

#### B. Results

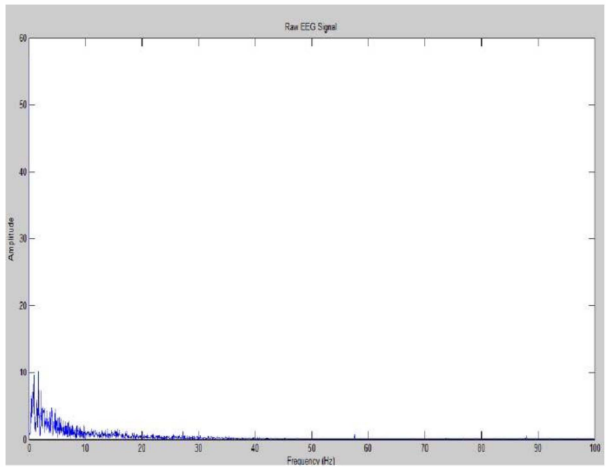

Fig.5. A sample of the raw EEG signal in the time domain.

Fig.5 shows an example of raw EEG data Physionet database. This example includes raw EEG activity measured by electrodes across several frontal sites.

The data of all the channels from each participant was transformed into frequency domain using FFT. The power of all the channels was computed by squaring the amplitude of the frequency domain signal. (Fig.6).

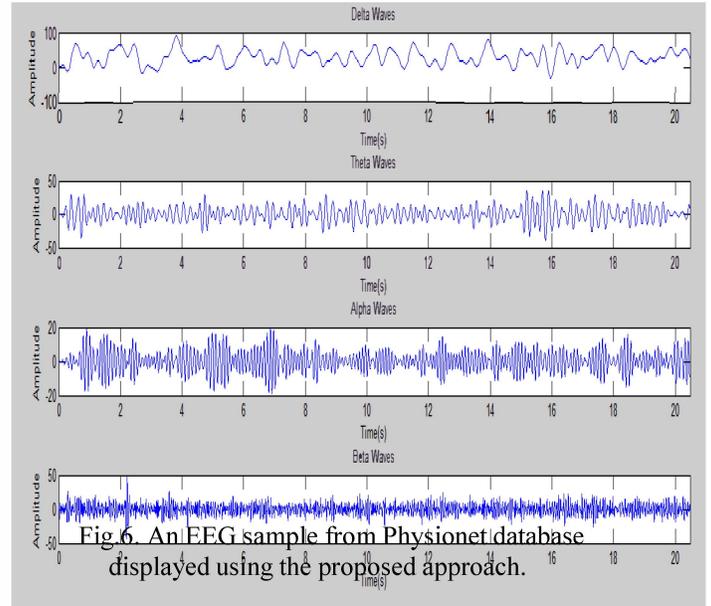

Fig.6. An EEG sample from Physionet database displayed using the proposed approach.

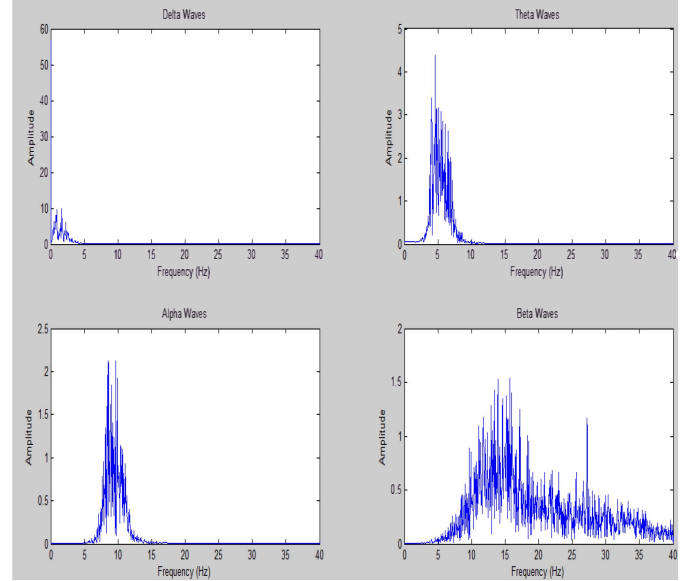

Fig.7. Delta, Theta, Alpha and Beta waves for one sample from Physionet database displayed using the proposed approach

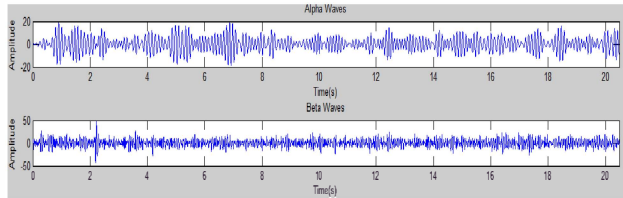

Fig.8. Alpha and Beta Waves during 20 s for one sample from Physionet database displayed using the proposed approach

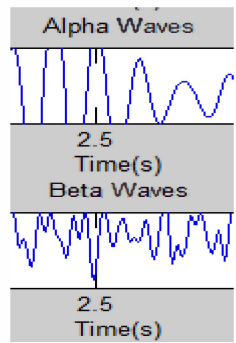

Fig.9. Comparison of Alpha and Beta bands waves in one moment for one sample from Physionet database displayed using the proposed approach.

The EEG power for delta, theta, alpha and bands are shown in Fig.7, Fig. 8 and Fig.9: The EEG power spectrum for all bands has been analyzed and an increase in alpha and theta power is observed when a subject is in transition from alert state to drowsy state. The increase is more significant in alpha band and less in theta band which is in agreement with the results in [13].

IV. CONCLUSION AND FUTURE WORKS

In this paper, we aim to develop an automatic system for drowsy driving identification or detection by analyzing EEG signals of the driver. This proposed has been validated in a standard database.

By evaluating the experimental results phase, we have achieved to get a degree of recognition advantageous to evaluate the drowsiness level; the degree of recognition is encouraged and suggestible to be ameliorated

As future researches, we are aspiring to suggest an advanced assistance system which monitors two types of risk: the inside vehicle dangers by evaluating the driver vigilance stage, and the outside vehicle dangers by detecting all different road moving and static objects during the vehicle works.